# First-Principles Understanding of the Electronic Band Structure of Copper-Antimony Halide Perovskite: The Effect of Magnetic Ordering

Xiaoming Wang[1,*], Weiwei Meng[1,2], Zewen Xiao[3], Jianbo Wang[2], David Mitzi[4], and Yanfa Yan[1,*]

[1] *Department of Physics and Astronomy, and Wright Center for Photovoltaic Innovation and Commercialization, The University of Toledo, Toledo, OH 43606, United States*

[2] *School of Physics and Technology, Center for Electron Microscopy, MOE Key Laboratory of Artificial Micro- and Nano-structures, and Institute for Advanced Studies, Wuhan University, Wuhan 430072, China*

[3] *Materials Research Center for Element Strategy, Tokyo Institute of Technology, Yokohama 226-8503, Japan.*

[4] *Department of Mechanical Engineering and Materials Science, and Department of Chemistry, Duke University, Box 90300 Hudson Hall, Durham, North Carolina 27708-0300, United State*

*e-mail address: wxiaom86@gmail.com, yanfa.yan@utoledo.edu*

**Abstract**: We report the understanding of the electronic band structure of $Cs_4CuSb_2Cl_{12}$ perovskite through first-principles density-functional theory calculations. We find that the most stable state has the antiferromagnetic configuration where each $[CuCl_6]$ octahedral chain along the [010] direction is antiferromagnetic. The reasonable band structure of the compound can be obtained only if both the correct magnetic order and the improved exchange interaction of the Cu $d$ electrons are taken into account.

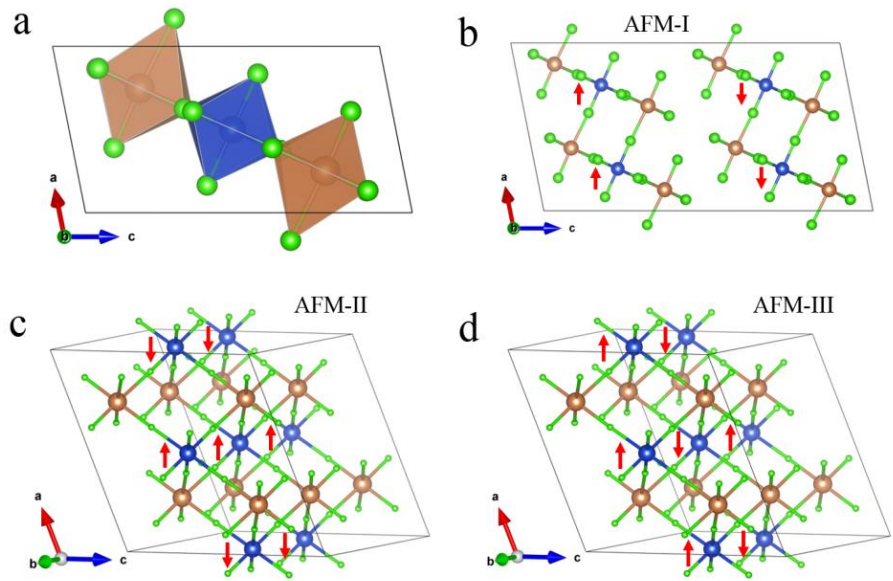

FIG. 1 (a) The primitive cell and (b)-(d) supercell with different antiferromagnetic configurations of $Cs_4CuSb_2Cl_{12}$. The Cu, Sb, and Cl atoms are denoted by the blue, brown, and green balls, respectively. The voids between the octahedrons are filled with Cs atoms, which are omitted for clarity. The red arrows denote the spin directions. AFM: antiferromagnetic.

Lead halide perovskites, used as solar cell absorbers, have attracted tremendous attention due to the rapid increase of the record high efficiency.[1–7] A great deal of efforts has been devoted to replacing the toxic lead atoms by proper non-toxic alternatives, resulting in the emergence of lead-free perovskites[8–15]. Very recently, $Cs_4CuSb_2Cl_{12}$ (CCSC) was synthesized with a reported direct band gap of 1.0 eV,[16] which is suitable for single-junction photovoltaic applications. CCSC has a layered structure with each Sb-Cu-Sb layer being composed of three sublayers with the Cu octahedral layer capped and corner-shared with two Sb octahedral layers, see Fig. 1a. The band structure was calculated via density-functional theory (DFT) with semi-local Perdew, Burke, and Ernzerhof[17] (PBE) functional in Ref. 16. The author claimed that a direct band gap of 0.98 eV (i.e. see Fig. 3 of Ref. 16) was obtained, consistent with their experiments. However, the Fermi level is located in the valence bands, indicating a metallic nature, which is not consistent with the semiconductor nature of the synthesized material. Typically, such metallic behavior of $Cu^{2+}$ containing compounds is due to the spin degeneracy of the Cu $d$ electrons in normal DFT.[18,19] Cu atom has the valence configuration of $4s^13d^{10}$. When forming $Cu^{2+}$-based compounds such as CCSC, the single $4s$ electron and one $3d$ electron lose, leading to an unpaired $d$ electron among the occupied $3d$ orbitals. This unpaired electron has two possible spin states, up and down. In CCSC, one of the spin states contributes to the bonding states, leaving the other in the antibonding states. Hence, the Cu $d$ band with this unpaired electron occupation is expected to split. However, in normal DFT calculations, the spin state is treated as degenerate. As a result, the band structure exhibits a metallic behavior.

To lift the band degeneracy, we conducted spin-polarized DFT calculations as implemented in the VASP code[20–22] with projector augmented-wave (PAW)[23] potentials. The experimental lattice parameters were used for the band structure calculations. We found that the magnetic ordering is crucial to get a reasonable band structure. The paramagnetic and ferromagnetic band structures were calculated through the primitive cell, as shown in Fig. 1a, with $a = b = 7.47$ Å, $c = 13.01$ Å, $\alpha = 70.96°, \beta = 109.04°, \gamma = 121.29°$. A Γ-centered $k$ grid of 4×4×2 was used to sample the Brillouin zone. Since there are strong correlations in $Cu^{2+}$ $d$ manifolds, we performed both PBE+$U$ ($U$ indicate on-site interaction) with a typical $U$ value of 8.5 eV[24,19,25] and HSE[26] calculations. As in the paramagnetic state, the electron spins are degenerate for all the three methods, resulting in the metallic behavior, as shown in Figs. 2a-c. For the ferromagnetic ordering, PBE failed to break the spin degeneracy due to the deficient exchange interaction. The band structure in Fig. 2d still indicates a metal. The local magnetic moment is zero everywhere. By improving the exchange interaction of the Cu $d$ electrons, both PBE+$U$ and HSE break the band degeneracy. A band gap opens between the two spin states. The band gap is indirect with 0.65 eV and 1.29 eV for PBE+U and HSE, respectively. The corresponding local magnetic moment of the Cu $d$ orbital is 0.57 and 0.60 $\mu_B$.

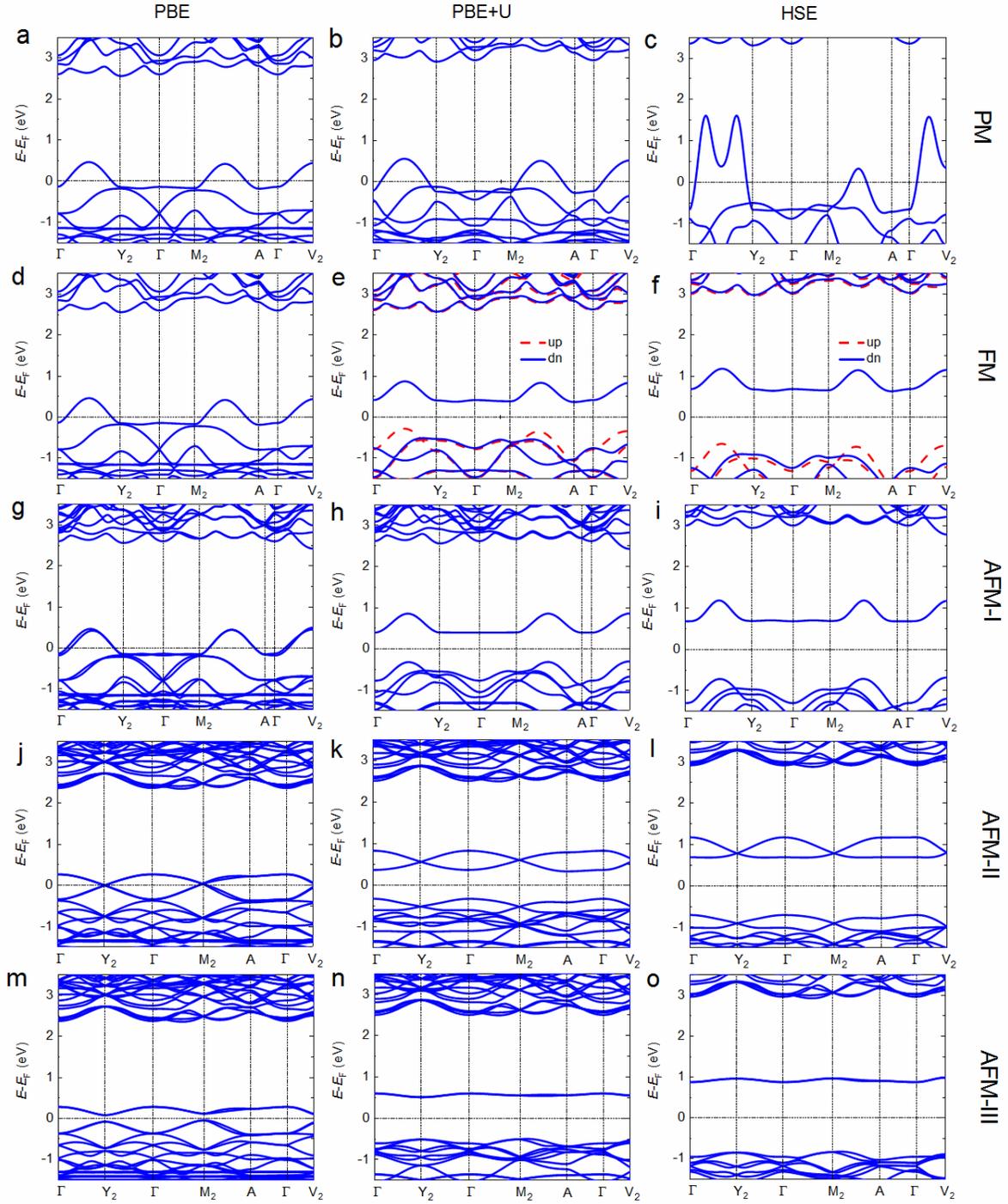

FIG. 2 Band structure of $Cs_4CuSb_2Cl_{12}$ with different magnetic configurations calculated by PBE, PBE+U, and HSE. PM: paramagnetic; FM: ferromagnetic; AFM: antiferromagnetic. Except for (e) and (f), the two spin states are almost degenerate, hence only one of them is shown for clarity.

There are three possible antiferromagnetic configurations for CCSC. First, each Sb-Cu-Sb layer can be ferromagnetic, but with antiferromagnetic coupling to its neighboring layers. We denote this configuration as AFM-I. A supercell with double size of the primitive cell in the $c$ axis was employed to evaluate the band structure (see Fig. 1b). Second, each Sb-Cu-Sb layer can also be antiferromagnetic. To this end, we constructed a supercell consisting of four units of $Cs_4CuSb_2Cl_{12}$ with $a = 13.03$ Å, $b = 14.65$ Å, $c = 13.01$ Å, $\alpha = \gamma = 90°$, $\beta = 111.98°$. With this supercell, there are two possible antiferromagnetic configurations, denoted as AFM-II and AFM-III, as shown in Figs. 1c and d, respectively. For the AFM-II antiferromagnetic configuration, each $[CuCl_6]$ octahedral chain along the [010] direction is ferromagnetic but antiferromagnetic with respect to its neighboring $[CuCl_6]$ octahedral chain, whereas for the AFM-III antiferromagnetic configuration, each $[CuCl_6]$ octahedral chain along the [010] direction is antiferromagnetic. The band structures of the antiferromagnetic states are shown in Fig. 2. For the PBE calculations, only the AFM-III configuration exhibits a semiconductor behavior with an indirect band gap of 0.13 eV. The local magnetic moment is 0.19 $\mu_B$. On the contrary, all the AFM states have an indirect band gap both for PBE+U and HSE calculations. Among the three AFM states, AFM-III has the largest band gap. The PBE+U band gap is 1.0 eV, in agreement with the experimental value. The local magnetic moment is 0.54 $\mu_B$. HSE predicts a larger band gap of 1.70 eV with a local magnetic moment of 0.58 $\mu_B$. Hence, HSE overestimates the band gap by 0.6 eV. HSE usually describes the band structures of $Cu^+$ compounds quite well, but gives a poor description of strongly correlated $Cu^{2+}$ $d$ states. An overestimation of the band gap by 1-1.3 eV was also found for CuO.[24] In this sense, PBE+U is superior to HSE for $Cu^{2+}$ compounds, but a Hubbard U is used as a parameter to fit the experimental results. However, the U value can also be determined from first-principles either by linear response[27] or constrained RPA calculations[28], which is beyond the scope of the present work.

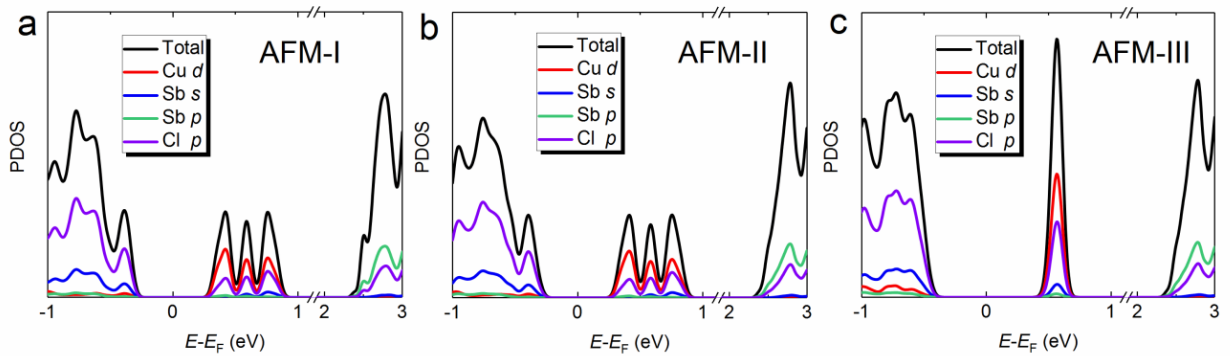

FIG. 3 PDOS of $Cs_4CuSb_2Cl_{12}$ in the antiferromagnetic configurations (a) AFM-I (b) AFM-II (c) AFM-III calculated from PBE+U.

The conduction band minimum (CBM) of all the AFM states are mainly contributed from Cu $d$ orbitals, as shown in Fig. 3, indicating the antibonding character of the Cu $d$ spin states. The optical absorption is dominated by Cl $p$ + Sb $s$ + Cu $d$ →

Cu $d$ + Cl $p$ transitions as in AFM-III. However, the Cu $d \rightarrow d$ transitions are dramatically reduced for AFM-I and AFM-II configurations, since their Cu $d$ orbital-derived conduction bands are more dispersive than that of AFM-III, as shown in Fig. 2. The separate flat Cu $d$ orbital-derived conduction band results in an isolated absorption peak in the calculated absorption coefficient spectrum for the AFM-III configuration (Figure 4). The relative energies and the band gaps of the all the magnetic configurations are summarized in Table I. For all the three methods, the AFM-III configuration was found the most stable state of $Cs_4CuSb_2Cl_{12}$ among the magnetic ordering configurations considered in this study.

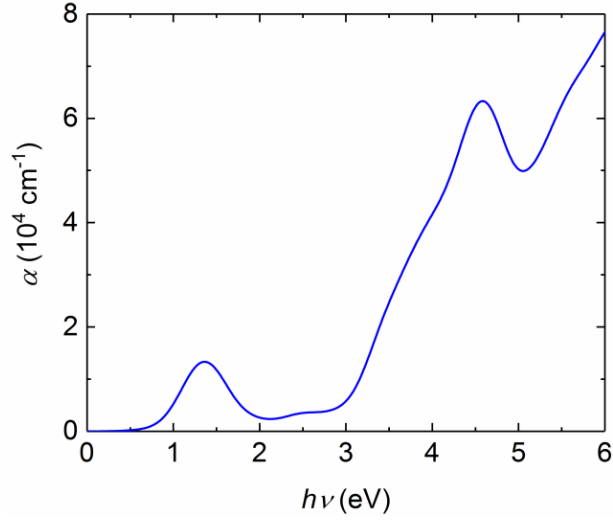

FIG. 4 Calculated optical absorption coefficient $\alpha$ by PBE+U for the the AFM-III configuration.

TABLE I The relative energy (meV)/band gap (eV)/local magnetic moment ($\mu_B$) on Cu $d$ orbitals of $Cs_4CuSb_2Cl_{12}$ for different magnetic configurations calculated by PBE, PBE+U, and HSE. PM: paramagnetic; FM: ferromagnetic; AFM: antiferromagnetic. The PM energy is taken as a reference.

| Method | PM | FM | AFM-I | AFM-II | AFM-III |
| --- | --- | --- | --- | --- | --- |
| PBE | 0.0/0.00/0.00 | 0.1/0.00/0.01 | -0.1/0.00/0.01 | -0.8/0.00/0.01 | -8.7/0.13/0.19 |
| PBE+U | 0.0/0.00/0.00 | -174.9/0.65/0.57 | -175.4/0.70/0.58 | -175.7/0.66/0.57 | -201.9/1.00/0.54 |
| HSE | 0.0/0.00/0.00 | -194.5/1.29/0.60 | -190.1/1.39/0.60 | -204.4/1.40/0.60 | -223.9/1.70/0.58 |

In conclusion, we studied the band structure of the newly synthesized copper-antimony halide perovskite $Cs_4CuSb_2Cl_{12}$ using first-principles DFT calculations. We found that the magnetic configuration is crucial to achieve a reasonable band structure. The most stable magnetic ordering of CCSC among those investigated in this study is determined to be antiferromagnetic AFM-III configuration, in which each [CuCl$_6$] octahedral chain along the [010] direction exhibits antiferromagnetic ordering. We employed PBE+U and HSE to improve the exchange interactions of the Cu $d$ electrons. The PBE+U band gap of 1.0 eV is in good agreement with the experimental value. HSE significantly overestimates the band gap by 0.6 eV, consistent with prior literature for another Cu$^{2+}$ compound, CuO.

**ACKNOWLEDGMENTS**

This work was funded in part by the Office of Energy Efficiency and Renewable Energy (EERE), U.S. Department of Energy, under Award Number DE-EE0006712, the Ohio Research Scholar Program, the National Science Foundation under contract no. CHE−1230246 and DMR−1534686, and the Office of Naval Research (Contract No. N00014-17-1-2223). This research used the resources of the National Energy Research Scientific Computing Center, which is supported by the Office of Science of the U.S. Department of Energy under Contract No. DE-AC02-05CH11231. The work at Wuhan University was supported by the National Natural Science Foundation of China (51671148, 51271134, J1210061, 11674251, 51501132, 51601132) and the Hubei Provincial Natural Science Foundation of China (2016CFB446, 2016CFB155).

*Disclaimer:* The information, data, or work presented herein was funded in part by an agency of the United States Government.  Neither the United States Government nor any agency thereof, nor any of their employees, makes any warranty, express or implied, or assumes any legal liability or responsibility for the accuracy, completeness, or usefulness of any information, apparatus, product, or process disclosed, or represents that its use would not infringe privately owned rights.  Reference herein to any specific commercial product, process, or service by trade name, trademark, manufacturer, or otherwise does not necessarily constitute or imply its endorsement, recommendation, or favoring by the United States Government or any agency thereof.  The views and opinions of authors expressed herein do not necessarily state or reflect those of the United States Government or any agency thereof.